\documentclass[12pt]{iopart}

\usepackage{iopams}  
\usepackage{graphicx}

\newcommand{\ket}[1]{| \, #1 \rangle}
\newcommand{\bra}[1]{ \langle #1 \,  |}

\begin{document}
\title{Direct observation of Hardy's paradox by joint weak measurement with an entangled photon pair}

\author{Kazuhiro Yokota, Takashi Yamamoto, Masato Koashi, and Nobuyuki Imoto}
\address{Department of Materials Engineering Science,
Graduate School of Engineering Science, Osaka University,
Toyonaka, Osaka 560-8531, Japan}
\address{CREST Photonic Quantum Information Project, 4-1-8 Honmachi, Kawaguchi, Saitama 331-0012, Japan}
\ead{yokota@qi.mp.es.osaka-u.ac.jp}

\date{\today}
\begin{abstract}
We implemented a joint weak measurement of the trajectories of two photons
in a photonic version of Hardy's experiment. The joint weak measurement has
been performed via an entangled meter state in polarization degrees of
freedom of the two photons. Unlike Hardy's original argument in which the
contradiction is inferred by retrodiction, our experiment reveals its
paradoxical nature as preposterous values actually read out from the meter.
Such a direct observation of
a paradox gives us new insights into the spooky action of quantum
mechanics.
\end{abstract}
\maketitle
\begin{figure}[t]
  \begin{center}
\includegraphics[scale=0.7]{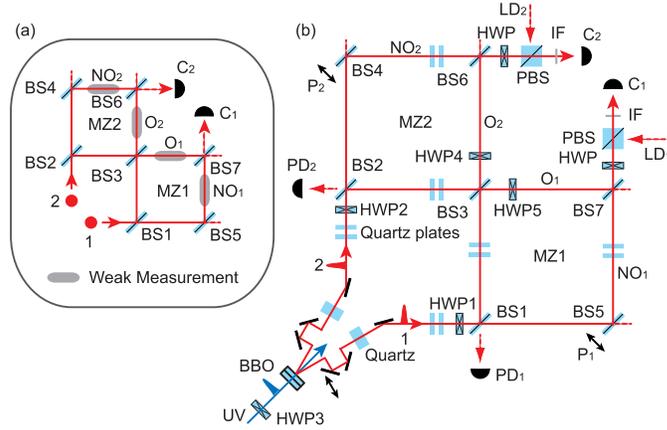}
  \end{center}
  \caption{Experimental setup for Hardy's paradox. (a) The photonic version
of Hardy's thought experiment. Each of the MZ interferometers MZ1
and MZ2 is composed of four 50:50 beam splitters (BS). (b) The schematics of
the experiment with joint weak measurement.  Entangled photon pairs are generated
via spontaneous parametric down-conversion from a pair of type I phase
matched 2mm thick BBO crystal \cite{PDC} pumped by a UV pulse (a central
wave length 395nm and an average power 180mW). The UV pulse is taken from the
frequency doubled Ti:sapphire laser (wavelength, 790nm; pulse width, 140fs;
repetition rate, 76MHz).
Quartz crystals are used to compensate the group velocity
mismatch and adjust the relative phase between horizontal and vertical
polarization state.
Polarization dependent phase shifts in MZs are
compensated by quartz plates in the MZs so that MZs give the same phase
shifts to any polarization state. Extra laser diodes (LDs) and photo
diodes (PDs) are used for adjusting and stabilizing the optical lengths of MZs via piezo stage
(Ps). The photons are detected by photon detectors (Cs).
The observed visibility of two photon interference at BS3 was 97.8 $\pm$ 0.3\% for horizontally polarized photons.}
  \label{fig:1}
\end{figure}

Although it is natural to ask what is the value of a physical quantity in the middle of a time evolution, it is difficult to answer such a question in quantum mechanics, especially when post-selection is involved.
Hardy's thought experiment \cite{Hardy} is a typical example in which
we encounter such a difficulty.
Figure \ref{fig:1}(a) shows a photonic version of the experiment,
which was demonstrated
by Irvine {\it et al.} recently \cite{Pho}.
The scheme consists of two Mach-Zehnder interferometers MZ1 and MZ2
with their inner arms ($O_1$, $O_2$) overlapping on each other at the 50:50 beam splitter BS3.
If photons 1 and 2 simultaneously arrive at BS3, due to a two-photon
interference effect, they always emerge at the same port. This
corresponds to the positron-electron annihilation in the original
thought experiment \cite{Hardy}.
The path lengths of MZ1 are adjusted so that photon 1 should
never appear at $C_1$ by destructive interference,
when photon 2 passes the outer arm
$NO_2$ and thus has no disturbance on MZ1.
The path lengths of MZ2 are adjusted similarly.
Then, a coincidence detection at $C_1$ and $C_2$ gives a paradoxical statement
on which paths the detected photons have taken.
The detection at $C_1$ ($C_2$) implies that MZ1 (MZ2) has been disturbed by photon 2
 (1) travelling along $O_2$ ($O_1$).
 We may thus infer that the
 conditional probabilities satisfy
\begin{eqnarray}
 P(O_1|C_1C_2)=P(O_2|C_1C_2)=1.
\label{eq:paradox1}
\end{eqnarray}
On the other hand, if both photons had taken the inner arms,
the coincidence detection would have never happened due to 
the two photon interference. Hence we infer that
\begin{eqnarray}
 P(O_1O_2|C_1C_2)=0.
\label{eq:paradox2}
\end{eqnarray}
The two inferred statements are apparently contradictory to
each other, which is the well-known Hardy's paradox.

One may argue that we should abandon the attempt to address 
the question itself, on the ground that the trajectory of photons
cannot be measured without utterly changing the time evolution.
But this reasoning is not necessarily true if we are allowed to 
repeat the same experiment many times.  
Aharonov {\it et al.} has proposed weak measurement \cite{W1, W2},
in which a measurement apparatus (meter) interacts with the system to be measured
so weakly that the state of the system is not significantly disturbed.
The readout of the meter from a single run of experiment may be subtle and
noisy, but by taking the average over many runs we can correctly estimate 
the expectation value of the measured observable,
$\bra{\psi} \hat{A}\ket{\psi}$, when the initial
state of the measured system is $\ket{\psi}$.
In this setup, we may ask what is the averaged readout over the runs 
in which the system is finally found to be in a state $\ket{\phi}$.
In the limit of no disturbance, this 
gives an operational way of defining what is the value of 
$\hat{A}$ at the middle of a time evolution from 
$\ket{\psi}$ to $\ket{\phi}$, and is found to be given by the real part of the 
following expression 
\begin{eqnarray}
	\hat{A}_{\bf w} \equiv \langle\phi|\hat{A}|\psi\rangle
/\langle\phi|\psi\rangle,   \label{eq:weak_def}
\end{eqnarray}
which is called the weak value of $\hat{A}$.
So far, related interesting features were discussed \cite{W-T1, W-T2,
W-T3, WH, W-T4, W-T6, W-T5} and experimental observations of weak values were 
reported \cite{W-E1, W-E2, Pryde, W-E4, W-E3}.

Suppose that weak measurements of trajectories are applied 
to Hardy's experiment at the shaded regions in Fig.~\ref{fig:1}(a).
The state of the photons entering these regions is 
$|\psi\rangle=(|NO_1\rangle|O_2\rangle+|O_1\rangle|NO_2\rangle+|NO_1\rangle|NO_2\rangle)/\sqrt{3}$,
and the coincidence detection retrodicts the state leaving the regions to
be 
$|\phi\rangle=(|NO_1\rangle-|O_1\rangle)(|NO_2\rangle-|O_2\rangle)/2$ \cite{WH}.
Then the weak values can be calculated to be  
\begin{eqnarray}
	&|O_1,O_2\rangle\langle O_1,O_2|_{\bf w}=0,  \ |NO_1,NO_2\rangle\langle NO_1,NO_2|_{\bf w}=-1, \nonumber \\
	&|O_1,NO_2\rangle\langle O_1,NO_2|_{\bf w}=1,  \ |NO_1,O_2\rangle\langle NO_1,O_2|_{\bf w}=1. \label{eq:joint_w}
\end{eqnarray}
The first equation implies Eq.~(\ref{eq:paradox2}) holds. We also see that
Eq.~(\ref{eq:paradox1}) holds since, for instance,
$|O_1\rangle\langle O_1|_{\bf w}=
|O_1,O_2\rangle\langle O_1,O_2|_{\bf w}+|O_1,NO_2\rangle\langle O_1,NO_2|_{\bf w}
=1$.
Hence the readout of the meter is indeed consistent with both
of the naively inferred conditions (\ref{eq:paradox1}) and 
(\ref{eq:paradox2}). The reason why these two contradictory conditions
are satisfied at the same time can now be ascribed to the appearance of 
a negative value, $|NO_1,NO_2\rangle\langle NO_1,NO_2|_{\bf w}=-1$.
It implies that the average readout over post-selected events falls 
on a value that never appears if no post-selection is involved.

In this paper, we report an experimental demonstration of weak
measurements applied to Hardy's experiment. Since the observables
to be measured are of path correlations between the two photons,
we need to implement a joint weak measurement that can extract such
correlations. Although a number of schemes for joint weak measurement
have been proposed \cite{Resch, Lundeen}, here we propose the most
intuitive approach using meter qubits initially prepared in an 
entangled state. We conducted measurements with varied strength, 
and confirmed that it works properly when each photon is injected
to a fixed arm. We also measured the visibility of the interferometers
to verify that the disturbance vanishes in the limit of weak 
measurement. Then we conducted Hardy's experiments and observed that,
as the measurement became weaker, three readouts moved toward 
the conditions (\ref{eq:paradox1}) and (\ref{eq:paradox2}) while 
another readout went to negative.

Let us first introduce our scheme of joint weak measurement for two
qubits in a general setting. 
Let $\{\ket{0},\ket{1}\}$ be the standard basis of a qubit.
For weak measurement of the observable 
$\ket{0}\bra{0}$ of a qubit, one can prepare another qubit 
in a suitable state and apply a controlled-NOT (C-NOT) gate between the two 
qubits \cite{Pryde}.
Here we want to carry out weak measurements of observables
$\ket{kl}_{s_1s_2}\bra{kl}$ $(k,l=0,1)$ for the two signal qubits, $s_1$ and $s_2$,
and we may use two meter qubits, $m_1$ and $m_2$, and apply two C-NOT gates in parallel.
The signal (meter) qubit corresponds to the control (target) qubit and the meter qubit is flipped when the signal qubit is in $|1\rangle$.
Note that in our implementation of Hardy's experiment, the photon path corresponds to the signal qubit and its polarization to the meter qubit.
We found that the desired weak measurements are achieved by a simple choice of the initial state of the meter qubits,
\begin{eqnarray}
|\xi\rangle_m &=& \delta|00\rangle_{m_1m_2}+\varepsilon(|01\rangle_{m_1m_2}+|10\rangle_{m_1m_2}+|11\rangle_{m_1m_2}) \nonumber \\ 
&=&(\delta-\varepsilon)|00\rangle_{m_1m_2}+\varepsilon\sum_{kl}|kl\rangle_{m_1m_2}, \label{eq:e_meter}
\end{eqnarray}
where $\delta^2+3\varepsilon^2=1$ and $\delta\ge \varepsilon\ge 0$. 
When the signal qubits are in $|\psi\rangle_s=\sum_{ij}c_{ij}|ij\rangle_{s_1s_2}$,
the application of the parallel C-NOT operation $U_{sm}$ results in
\begin{eqnarray}
& &U_{sm}|\psi\rangle_s|\xi\rangle_m \nonumber \\
&=&\sum_{ij}c_{ij}|ij\rangle_{s_1s_2}[(\delta-\varepsilon)|ij\rangle_{m_1m_2}+\varepsilon\sum_{kl}|kl\rangle_{m_1m_2}] \nonumber \\
&=&\sum_{kl}[(\delta-\varepsilon)|kl\rangle_{s_1s_2}\langle kl|+\varepsilon]|\psi\rangle_s|kl\rangle_{m_1m_2}. \label{eq:cnot}
\end{eqnarray}
The meter qubits will then be measured in the basis
$\{\ket{kl}_{m_1m_2}\}$
to produce an outcome $(k,l)$.
If the signal qubits are in state  $|ij\rangle_{s_1s_2}$ initially, 
the probability $P_m(k,l)$ of outcome $(k,l)$ becomes 
\begin{eqnarray}
{P}_m(k,l)=\left\{
\begin{array}{ll}
\delta^2 & ((k,l)=(i,j))\\
\varepsilon^2 & ((k,l)\neq (i,j)).
\end{array}
\right.
\end{eqnarray}
We see that the outcome $(i,j)$, corresponding to the state of the
signal, has a larger probability than the others. The contrast 
$\delta^2-\varepsilon^2$ will be regarded as the measurement strength.
When $\delta^2-\varepsilon^2=1$ $(\delta=1, \varepsilon=0)$, this scheme gives a projection 
measurement (strong measurement)
on the signal qubits. Decreasing $\delta^2-\varepsilon^2$ makes the
measurement weaker, and at $\delta^2-\varepsilon^2=0$
 $(\delta=\varepsilon=1/2)$, the operation of $U_{sm}$ introduces no
disturbance on the signal qubits (Appendix A).
Let us introduce a normalized readout $R(k,l)$ by 
\begin{eqnarray}
R(k,l) &\equiv & [P_m(k,l)-\varepsilon^2]/(\delta^2-\varepsilon^2). 
\end{eqnarray}
Then, regardless of the value of $\delta^2-\varepsilon^2>0$, 
we have $R(k,l)=1$ for $(k,l)=(i,j)$ and $R(k,l)=0$ otherwise.
When the initial state of the signal qubits is $\ket{\psi}_s=\sum_{ij}c_{ij}|ij\rangle_{s_1s_2}$,
the normalized readout becomes $R(k,l)=|c_{k,l}|^2$, 
which coincides with the expectation value 
$\langle \ket{kl}_{s_1s_2}\bra{kl} \rangle$. Of course, for the estimation 
of $R(k,l)$ and hence of $P_m(k,l)$, we need to repeat the 
preparation of the initial state $\ket{\psi}_s$ 
and the measurement many times.

Now in a situation like Hardy's experiment, we are interested in the
readout of the meter on condition that the signal qubits are finally 
measured to be in state $\ket{\phi}_s$.
From the definition (\ref{eq:weak_def}), we have
\
\begin{eqnarray}
& &_s\langle \phi|U_{sm}|\psi\rangle_s|\xi\rangle_m \nonumber \\
&=&\sum_{kl}[(\delta-\varepsilon)|kl\rangle_{s_1s_2}\langle kl|_{\bf w}+\varepsilon]_s\langle \phi|\psi\rangle_s|kl\rangle_m.
\end{eqnarray}
The conditional probability of the outcome $(k,l)$ is then given by
\begin{eqnarray}
{P}_m(k,l|\phi)&=&|(\delta-\varepsilon)|kl\rangle_{s_1s_2}\langle kl|_{\rm w}
+\varepsilon|^2 /[1-(\delta-\varepsilon)^2\zeta ]
\end{eqnarray}
with $\zeta \equiv 1-\sum_{ij}||ij\rangle_{s_1s_2}\langle ij|_{\rm w}|^2$.
The normalized readout is found to be
\begin{eqnarray}
& &R(k,l|\phi)
\equiv [{P}_m(k,l|\phi)-\varepsilon^2]/(\delta^2-\varepsilon^2) \label{eq:ij} \\
&=&\frac{2\varepsilon(\delta-\varepsilon){\rm Re}
[|kl\rangle_{s_1s_2}\langle kl|_{\rm w}]+(\delta-\varepsilon)^2
[||kl\rangle_{s_1s_2}\langle kl|_{\rm w}|^2+\varepsilon^2 \zeta]}
{(\delta^2-\varepsilon^2)[1-(\delta-\varepsilon)^2 \zeta]} \nonumber \\
&\to & {\rm Re}[|kl\rangle_{s_1s_2}\langle kl|_{\rm w}]
\ \ \ (\delta^2-\varepsilon ^2 \to 0). \label{eq:j_weak}
\end{eqnarray}
We see that the normalized readout in the limit of no disturbance 
coincides with the real part of the weak value.

In our experiment, the signal qubit corresponds to the paths taken by the photon
as $\ket{0}=\ket{NO}$ and $\ket{1}=\ket{O}$, whereas the meter qubit is
assigned to the polarization of the same photon as 
$\ket{0}=\ket{+}$ and $\ket{1}=\ket{-}$, where
 $|\pm\rangle \equiv(|H\rangle\pm |V\rangle)/\sqrt{2}$ and $H/V$ is the
 horizontal/vertical polarization. The interferometers in 
Fig.~\ref{fig:1}(a) are constructed so that there is no 
polarization dependence, except for the shaded region where 
wave plates are placed to realize $U_{sm}$. 
The polarization dependence of the coincidence events is then analyzed 
to determine $P_m(k,l|\phi)$.

\begin{figure}[tb]
  \begin{center}
	 \includegraphics[scale=0.6]{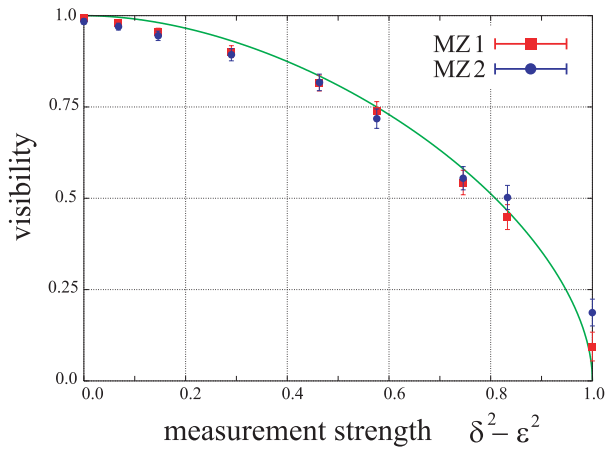}
  \end{center}
  \caption{Relation between the visibility of MZ1(2) and the measurement strength. The solid curve represents the theoretically expected visibility, $2\varepsilon(\delta+\varepsilon)$.}
  \label{fig:2}
\end{figure}

The detail of the setup is shown in Fig.~\ref{fig:1}(b). We generate the photon pairs in a nonmaximally entangled state $\eta |HH\rangle_{12}+\bar{\eta}|VV\rangle_{12}$ from PDC, where $\eta^2+\bar{\eta}^2=1$ and $\eta$ and $\bar{\eta}$ are properly adjusted to be real numbers. The coefficients $\eta$ and $\bar{\eta}$,
which can be tuned by HWP3, are determined by the ratio between the coincidence counts of $|HH\rangle$ and $|VV\rangle$. 
The state is transformed into $|\xi\rangle_m$ by rotating the polarization of two photons by half wave plates HWP1 and HWP2. The measurement strength is simply calculated by the measured $\eta$ and $\bar{\eta}$ and the angle of HWP1 and HWP2. 
After the photons pass through BS1, BS2 and BS3, the two-photon state is represented as $|\psi\rangle_{s}|\xi\rangle_m$. The C-NOT operations between signals and meters are simply performed by flipping the polarization as $|\pm\rangle \to |\mp\rangle$ via HWP4 and HWP5 in the overlapped arms, $O_1$ and $O_2$.  
The polarization basis ($\ket{+}, \ket{-}$) for the meter measurement is selected by HWP and PBS just before detectors $C_1$ and $C_2$. 
The coincidence counts, typically a few thousands, are accumulated over 5-12 min per basis to determine each $R(k,l|\phi)$.
All error bars assume the Poisson statistics of the counts.

First we show the observed trade-off between the measurement strength and the visibilities of interference at MZ1 and MZ2. The measurements were done as follows: We blocked the arm $O_2$($O_1$) and recorded the coincidence detection at $C_1$ and $C_2$ by changing the path length via piezo stage P$_1$(P$_2$). In order to measure the polarization-independent visibilities, we measured the interference fringes in the states $|++\rangle$, $|+-\rangle$, $|-+\rangle$, and $|--\rangle$ independently and calculated the average visibility.
The visibilities for various measurement strengths are shown in Fig.~\ref{fig:2}. We can clearly see that weakening the measurement makes the visibility larger (Appendix B).

\begin{figure*}[t]
  \begin{center}
	\includegraphics[scale=0.8]{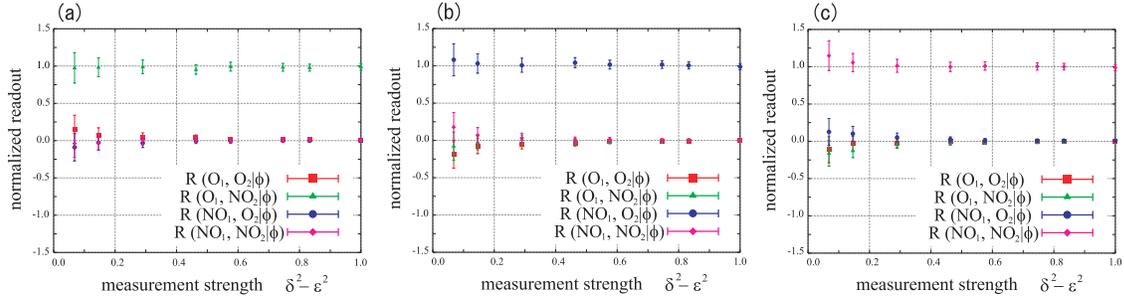}
  \end{center}
  \caption{Experimental results when the initial states are in 
(a)$|O_1,NO_2\rangle$, (b)$|NO_1,O_2\rangle$, and (c)$|NO_1,NO_2\rangle$.}
  \label{fig:3}
\end{figure*}

We also show that our measurement gives proper values of the readouts
$R(k,l|\phi) $ when the photons travel the fixed arms $|O_1,NO_2\rangle,
|NO_1,O_2\rangle,$ and $|NO_1,NO_2\rangle$. We blocked the arms of no
interest and determined $R(k,l|\phi) $ from the coincidence detection on the
basis \{ $|++\rangle$, $|+-\rangle$, $|-+\rangle$, $|--\rangle$ \} . The
experimental results are plotted in Fig.~\ref{fig:3}. In any measurement
strength, the corresponding readout for the signal state is close to 1
and the other readouts are close to 0. This clearly shows the apparatus
properly measures the trajectories of the photons. 
The systematic deviation from the expected values at small values of
$\delta^2-\varepsilon ^2$ is believed to be the results of very tiny
misalignments of the axes of the wave plates.
Such an enhancement is a feature of weak measurement and can be used for
super-sensitive measurements as shown in \cite{W-E3}.

\begin{figure}[t]
  \begin{center}
	\includegraphics[scale=0.6]{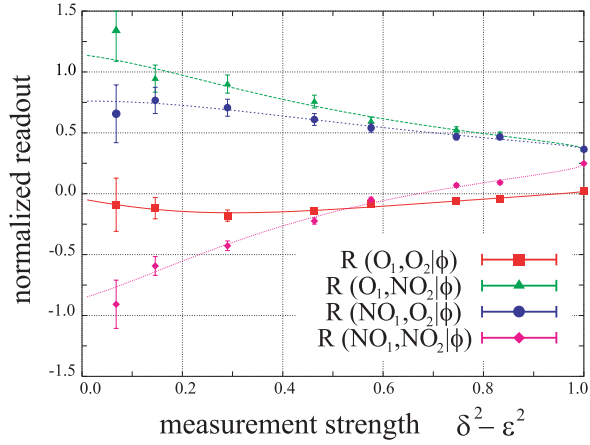}
  \end{center}
  \caption{Observed readouts in Hardy's experiment.
The curves that roughly fit the data points are of a simple model of imperfections, and are only shown to guide the eye.}
  \label{fig:4}
\end{figure}

With this meter, we performed measurements of the trajectories of the
photons in Hardy's thought experiment. The observed readouts
are shown in Fig.~\ref{fig:4}.
Let us see the results around 0.3 measurement strength.
The visibility above 0.9 shown in Fig.~\ref{fig:2}
suggests that the measurement is fairly weak,
and the systematic deviation in Fig.~\ref{fig:3} is small
at this strength. For each of the photons, the readouts suggest
a high probability of taking the overlapped arm:
$R(O_1|\phi)\equiv R(O_1,NO_2|\phi)+R(O_1,O_2|\phi)=0.72\pm 0.09$
and $R(O_2|\phi)=0.53\pm 0.09$.
Nonetheless, the readout for the joint probability is
around zero, namely, $R(O_1,O_2|\phi)=-0.18\pm 0.05$.
We also see that $R(NO_1,NO_2|\phi)=-0.43\pm 0.04$ takes a
large negative value, which cannot be interpreted as a probability.
Although the statistical and systematic errors are large around zero
measurement strength, we see that
$R(O_1,NO_2|\phi) $, $R(NO_1,O_2|\phi) $, $R(O_1,O_2|\phi) $, and
$R(NO_1,NO_2|\phi)$ approach the expected weak values in Eq.~(\ref{eq:joint_w})
when the measurement strength goes to 0.

Let us emphasize here that our experimental results by themselves can elucidate the
paradoxical nature of Hardy's experiment, without any reference to the
theory explaining how our measurement works. From the results in Fig.~\ref{fig:3}, we
see that our measurement apparatus should properly give the probabilities on
the trajectories of the photons, if such quantities ever exist. Then, in
Hardy's setup, the same measurement presents a contradictory statement 
in Fig.~\ref{fig:4} that the probability of finding a photon in arm $O_1$ approaches
1 and so does the probability for arm $O_2$, whereas the joint probability of
photons being in both of the arms stays at about 0. Moreover, the readout
points to a negative value for the joint probability for arms $NO_1$ and
$NO_2$. Unlike Hardy's original argument, our demonstration reveals the
paradox by observation, rather than inference.

We have experimentally demonstrated a joint weak measurement with an
entangled photon pair and directly observed paradoxical results in Hardy's
thought experiment. Our demonstration clearly reproduces Hardy's paradox
in an experimentally accessible manner. Weak measurements have attracted
attention due to the possibility of achieving small-backaction and
high-sensitivity measurements by simple optical setups. We believe the
demonstrated joint weak measurement is useful not only for exploiting
fundamental quantum physics, but also for various applications such as
quantum metrology and quantum information technology.

Note added. During the preparation of this manuscript, we came to know of a
work by Lundeen and Steinberg \cite{L_EH}, which has demonstrated similar results by
using, interestingly, different methods for joint weak measurement
and for constructing Hardy's interferometer.

\section*{Acknowledgment}
This work was partially supported by the JSPS (Grant-in-Aid for
Scientific Research (C) number 20540389) and by the MEXT (Grant-
in-Aid for Scientific Research on Innovative Areas number 20104003,
Global COE Program and Young Scientists(B) number 20740232).

\appendix
\section{}
The positive operator valued measure (POVM) for the measurement on the signal qubits is given by
\begin{eqnarray}
\hat{\Pi}_{kl}&=&\delta^2|kl\rangle_{s_1s_2}\langle kl|+\varepsilon^2\sum_{ij\neq kl} |ij\rangle_{s_1s_2}\langle ij| \nonumber \\
&=&(\delta^2-\varepsilon^2)|kl\rangle_{s_1s_2}\langle kl|+\varepsilon^2.
\end{eqnarray}
The same POVM can be realized also by a nonentangled initial state of the meter qubits, $\delta^2|00\rangle_{m_1m_2}\langle 00|+ \varepsilon^2(|01\rangle_{m_1m_2}\langle 01|+|10\rangle_{m_1m_2}\langle 10|+|11\rangle_{m_1m_2}\langle 11|)$. This separable state of the meter qubits, however, fails to derive a joint weak value, because the signal qubits are utterly disturbed even if $\delta^2-\varepsilon^2=0$.

\section{}
A general relation between the visibility $V$ and the measurement
strength was discussed in \cite{Englert}. It was shown that $V^2+K^2\le 1$
holds, where $K\equiv 1-2p_{\rm err}$ stands for the measurement strength
defined in terms of the error probability $p_{\rm err}$ in the
which-path measurement. In our case, the initial marginal
state of the meter qubit $m_1$ is a mixed state, which is,
from Eq.~(\ref{eq:e_meter}), calculated to be a
mixture of state
$\ket{\phi_0}_{m_1}\equiv
(\delta\ket{0}_{m_1}+\varepsilon\ket{1}_{m_1})/\sqrt{p_0}$
with probability $p_0\equiv \delta^2+\varepsilon^2$ and
$\ket{\phi_1}_{m_1}\equiv (\ket{0}_{m_1}+\ket{1}_{m_1})/\sqrt{2}$
with $p_1\equiv 2\varepsilon^2$. The state $\ket{\phi_0}_{m_1}$ leads
to a measurement strength
$K_0=(\delta^2-\varepsilon^2)/(\delta^2+\varepsilon^2)$
and visibility $V_0=\sqrt{1-K_0^2}=2\varepsilon
\delta/(\delta^2+\varepsilon^2)$, whereas
$\ket{\phi_1}_{m_1}$ leads to $K_1=0$ and $V_1=1$.
Hence, on average, the measurement strength becomes
$K_{\rm ave}\equiv p_0K_0+p_1K_1=\delta^2-\varepsilon^2$
and $V_{\rm ave}\equiv p_0V_0+p_1V_1=2\varepsilon(\delta+\varepsilon)$,
which is shown as the solid curve in Fig. 3. Due to the mixture,
the averaged values $K_{\rm ave}$ and $V_{\rm ave}$ do not saturate
the general bound $V_{\rm ave}^2+K_{\rm ave}^2\le 1$.

\end{document}